# Multiscale Decoupling in Peptide-Membrane Interactions: From Elastic Softening to Domain-Mediated Stiffening

Veerendra K. Sharma[1, 2*], Shuo Qian[3*], Gergely Nagy[3], Sonam Raghav[4], Sajal K. Ghosh[4]

[1]*Solid State Physics Division, Bhabha Atomic Research Centre, Mumbai, India 400 085*

[2]*Homi Bhabha National Institute, Mumbai, India 400 094*

[3]*Neutron Scattering Division, Oak Ridge National Laboratory, Oak Ridge, TN 37830, United States*

[4] *Department of Physics, School of Natural Sciences, Shiv Nadar Institution of Eminence, Gautam Buddha Nagar, Uttar Pradesh 201314, India*

## Abstract

Membrane composition plays a central role in determining how antimicrobial peptides (AMPs) interact with and alter the physical properties of lipid membranes. Here, we investigate the composition-dependent mechanical and structural response of lipid membranes to the AMP aurein using neutron spin-echo (NSE) spectroscopy, neutron membrane diffraction (NMD), and pressure ($\pi$)-area ($A$) isotherms, complemented by previously reported quasielastic neutron scattering (QENS) measurements. In a zwitterionic DMPC membrane, aurein induces progressive softening, manifested by a concentration-dependent decrease in both the bending rigidity and area compressibility modulus. NMD reveals a slight bilayer thinning but no prominent peptide-associated contribution within the deuterated hydrocarbon region, indicating that the progressive mechanical softening is not accompanied by increasingly deep peptide penetration into the hydrophobic core. In contrast, anionic phosphatidylglycerol (PG)-containing DMPC/DMPG membranes exhibit a non-monotonic response, with initial softening followed by pronounced stiffening at higher peptide concentrations. Complementary $\pi$-$A$ isotherm measurements independently reproduce these contrasting mechanical responses, showing progressive softening of DMPC but stiffening of DMPC/DMPG. Most importantly, the combined NSE and QENS results reveal a striking multiscale decoupling: aurein suppresses molecular-scale lipid lateral diffusion while simultaneously reducing collective membrane stiffness at the mesoscopic scale, demonstrating that molecular lipid dynamics and collective membrane mechanics can evolve in opposite directions under the same peptide perturbation. Together, these results show that membrane composition governs the collective mechanical response to aurein, while molecular lipid dynamics can respond independently at shorter length and time scales. This work provides a physical framework for understanding composition-dependent peptide-membrane interactions and establishes a testable basis for extending such studies to more complex membrane environments.

***Corresponding Authors**: sharmavk@barc.gov.in (VKS); qians@ornl.gov (SQ)

## INTRODUCTION

Antibiotic resistance is an escalating global health threat, rendering many conventional treatments ineffective and fueling the rise of multidrug-resistant 'superbugs'[1]. There is a pressing need to find novel antibacterial agents to combat this problem. Antimicrobial peptides (AMPs) have attracted significant attention as promising alternatives due to their broad-spectrum activity, rapid mode of action, and low propensity for inducing resistance[2], [3], [4], [5]. As key components of the innate immune system, AMPs provide an evolutionarily conserved defense mechanism against a wide range of pathogens [6]. However, realizing their full therapeutic potential requires a detailed understanding of their molecular mechanisms, particularly the physical principles governing their interaction with cellular membranes.

A central feature of AMP activity is their ability to selectively target bacterial membranes while sparing mammalian cells. This selectivity arises primarily from differences in membrane composition: bacterial membranes are enriched in anionic lipids such as phosphatidylglycerol (PG), whereas the outer leaflet of mammalian membranes is predominantly composed of zwitterionic lipids such as phosphatidylcholine (PC). Consequently, AMP-membrane interactions are governed by a delicate interplay of electrostatic attraction and hydrophobic insertion[7], [8], [9], [10], [11], [12], [13], [14]. Beyond simple binding, these interactions modulate membrane structure and mechanics, which ultimately dictate membrane destabilization pathways. Understanding how lipid composition regulates these emergent mechanical responses is therefore essential for establishing a physical basis of AMP selectivity.

Aurein peptides secreted from the skin of Australian Litoria frogs, represent a class of short, cationic AMPs. Among them, Aurein 1.2 (hereafter referred to as aurein) is one of the shortest members, consisting of just 13 amino acids with a C-terminal amidation (Figure-S1). Despite its small size, aurein exhibits potent antimicrobial activity[15]. It also shows broad-spectrum antibiotic and anticancer properties and shares sequence similarity with segments of human AMPs such as LL-37. In aqueous environments, aurein remains unstructured; however, it adopts α-helical conformation upon interaction with lipid membranes[16], [17]. Unlike pore-forming peptides such as alamethicin or melittin, aurein does not form stable transmembrane pores even at high peptide-to-lipid ratios[16]. Instead, it perturbs membranes through subtle yet functionally significant mechanisms, including modulation of lipid packing, membrane thinning, alteration of membrane mechanical properties (e.g., bending rigidity), and peptide-induced lateral organization[16], [17], [18]. These features make aurein an ideal model system to investigate how short AMPs regulate membrane mechanics without inducing catastrophic structural disruption.

Neutron scattering techniques provide a unique and powerful framework to probe membrane structure and dynamics across multiple length and time scales[19], [20]. In particular, neutron membrane diffraction (NMD) enables direct reconstruction of the scattering length density (SLD) profile across the bilayer normal, offering Å-scale resolution of membrane architecture, including headgroup organization, bilayer thickness, and peptide localization within the membrane[19], [21], [22]. Complementarily, neutron spin-echo (NSE) spectroscopy probes collective membrane dynamics at mesoscopic length scales, providing direct access to membrane bending fluctuations and allowing quantitative determination of key elastic parameters such as the bending modulus (κ) and area compressibility modulus ($K_A$) [20], [23],[24], [25], [26], [27]. The combination of NMD and NSE is especially powerful, as it links structural organization with emergent mechanical properties, enabling a comprehensive understanding of how peptide binding modulates membrane architecture and elasticity across scales. Recently, NMD and NSE were used to investigate aurein's interaction with mixed, anionic–zwitterionic lipid membranes[18]. NMD results showed that even at the highest studied peptide to lipid (P/L) ratios, aurein primarily localizes at the interface between the lipid headgroups and acyl chains. NSE measurements revealed that aurein modulates membrane mechanical properties in a concentration-dependent manner: it softens membranes at low concentrations but induces increased rigidity at higher concentrations, highlighting its dual mechanical role in membrane remodeling.

While interactions with charged and mixed membranes are central to antimicrobial peptide activity, understanding peptide interactions with purely zwitterionic membranes is equally important for isolating the contribution of membrane composition under well-defined conditions. Zwitterionic phosphatidylcholine membranes provide a controlled platform in which strong electrostatic attraction between cationic peptides and anionic lipids is absent, allowing the role of hydrophobic and other non-electrostatic interactions to be examined more directly. A comparison between zwitterionic and anionic-containing membranes can therefore reveal how lipid composition influences peptide association, lateral organization, and emergent membrane mechanics. Such a controlled comparison provides mechanistic insight into how membrane composition modulates the physical response to peptide, without implying that either model directly reproduces the complexity of native biological membranes.

In this work, we investigate the structural and mechanical response of zwitterionic 1,2-dimyristoyl-sn-glycero-3-phosphocholine (DMPC) bilayers to aurein using NSE spectroscopy, NMD and pressure-area ($\pi$-$A$) isotherm. By integrating the present measurements with our previously reported quasielastic neutron scattering (QENS) results[17] and benchmarking them against NSE study on anionic DMPC/DMPG membranes[18], we establish a comparative framework that isolates the membrane-level effects of lipid composition on aurein-membrane interactions under well-defined model conditions. Throughout this work, P/L denotes the nominal peptide-to-lipid molar ratio, as the controlled experimental variable, not the membrane-bound peptide concentration or activity, which may depend on lipid composition. We find that aurein induces continuous, concentration-dependent softening of zwitterionic DMPC membranes, manifested by reductions in both bending rigidity and area compressibility modulus, without the high-concentration stiffening observed in anionic PG-containing membranes[18]. Importantly, the combined NSE and QENS results reveal a striking multiscale decoupling: aurein suppresses molecular-scale lipid lateral diffusion while simultaneously reducing collective membrane stiffness at the mesoscopic scale. Thus, molecular lipid dynamics and collective membrane elasticity can exhibit contrasting responses to the same peptide perturbation, reflecting distinct manifestations of peptide-membrane interactions at different length and time scales. Together, these results demonstrate that lipid composition strongly influences the mechanical pathway of aurein-induced membrane remodeling and provide a testable physical framework for investigating AMP–membrane interactions in increasingly complex membrane environments.

## EXPERIMENTAL DETAILS

### Materials

Tail deuterated DMPC (d54- DMPC) and fully protonated DMPC were purchased from Avanti Polar Lipids, Inc. (Alabaster, AL, U.S.A.). Aurein peptide (GLFDIIKKIAESF-$NH_2$, > 95% purity) was synthesized by GenScript USA Inc. (Piscataway, NJ, USA) and $D_2O$ was obtained from Sigma-Aldrich. All materials were used as received without further purification

**Preparation of Substrate-Supported Membranes**

Substrate-supported lipid multilayers for NMD experiments were prepared by mixing d54-DMPC with aurein at P/L ratios of 0, 1.25, 2.5, and 4 mol%. The lipid-peptide mixtures were dissolved in a 1:1 (v/v) trifluoroethanol/chloroform solution to ensure homogeneous mixing, and deposited onto clean quartz substrates (19 mm × 38 mm; GM Quartz, Oakland, CA, USA). The samples were dried under a gentle stream of air followed by vacuum drying to remove residual organic solvents. The resulting films were hydrated in a sealed chamber containing saturated $D_2O$ vapor at 30 °C for at least 12 h to achieve well-aligned multilamellar structures. For each condition, two substrates containing approximately 5 mg of lipid each were stacked to enhance the diffraction signal.

**Preparation of Unilamellar Vesicles**

DMPC unilamellar vesicles (ULVs) were prepared by extrusion method as described previously[8], [9], [28] [29], [30]. Briefly, DMPC was dissolved in chloroform, and the solvent was evaporated under a gentle stream of dry nitrogen to form a thin lipid film. Residual chloroform was removed by vacuum drying overnight. The lipid film was then hydrated with $D_2O$ at 310 K and subjected to three freeze-thaw cycles by alternating between a warm water bath (320 K) and a freezer (253 K) to ensure uniform dispersion. ULVs were obtained by extruding the lipid suspension through a polycarbonate membrane (pore diameter ~100 nm) using a mini-extruder (Avanti Polar Lipids) for at least 31 passes. Previous studies have shown that residual multilamellarity in extruded vesicles can be reduced through several complementary approaches [29,30]. We incorporated several of these strategies (using pure water without any salt to minimize the ionic strength of the aqueous medium, and increasing extrusion cycles to 31 through smaller pore diameter (~100 nm) polycarbonate membrane ) in our vesicle preparation method so that we can minimize the residual multilamellar structures. A stock solution of peptide 10 mg/ml aurein was prepared in 100 % $D_2O$. For NSE measurements, samples were prepared at a lipid concentrations of 20 mg/ml with aurein concentrations of 0, 1.25, 2.5 and 4 mol % i.e. (P/L=0, 1/80, 1/40 and 1/25). SANS measurements of selected DMPC-aurein samples had been performed on the Bio-SANS beamline of the High Flux Isotope Reactor (HFIR) of the Oak Ridge National Laboratory (ORNL) [31], and the resulting scattering profiles showed no distinct signatures indicative of a significant MLV contribution [Fig S2].

**Neutron Membrane Diffraction (NMD)**

NMD measurements were performed on EQ-SANS instrument at the Spallation Neutron Source (SNS) of ORNL[32]. Samples were mounted in a sealed sample cell in grazing-incidence geometry relative to the neutron beam and maintained at 30 °C under controlled relative humidity (RH)[21]. The RH was set to 94 % and 98% using a humidity generator (Model RH-200, L&C Science, USA) supplied with a reservoir containing 8% (v/v) $D_2O$. The 8% (*v/v*) $D_2O$ contrast was selected for two important reasons: (1) The NSLD of 8% (v/v) $D_2O$ is zero, minimizing the contribution of water in the reconstructed NSLD profiles for data interpretation; (2) this ensures diffractions measured at different RH sample a common continuous transform of the bilayer structure, enabling better phase determination by the swelling method [33]. The instrument was configured with a neutron wavelength band of 2.5–6.8 Å selected by the chopper system and a sample-to-detector distance of 1.3 m. Diffraction patterns were collected after equilibration at the target RH. An oscillating rocking scan was performed by rotating the sample from $\omega = 0°$ to 14° in steps of 0.1°.

**Neutron Spin Echo (NSE) Spectroscopy**

NSE measurements were carried out on 2 (*w/w*) % DMPC ULVs with varying concentrations of aurein at physiological temperature 37°C using the NSE spectrometer (BL-15)[34] at the SNS, ORNL. Temperature was controlled using an oil circulation bath with an accuracy of ±0.1 °C. The samples were loaded into Hellma quartz cells (4 cm ×4 cm) with a thickness of 3 mm. For the measured incoherent NSE data, corrections were performed using resolution data from a grafoil sample and background from $D_2O$. The data reduction was performed with the standard ECHODET software package of the SNS-NSE instrument.

**Surface Pressure-Area Isotherms**

Surface pressure-area isotherms were recorded using a medium-sized (36.4 × 7.5 × 0.4 $cm^3$) Langmuir–Blodgett (LB) trough (KSV NIMA, Biolin Scientific) equipped with a platinum Wilhelmy plate and movable Delrin barriers. DMPC was dissolved in a chloroform: methanol solvent mixture at a volume ratio of 3:1. Aurein was dissolved separately in the same solvent mixture. The two solutions were subsequently mixed at the required lipid-to-peptide molar ratio (mol%) to ensure homogeneous mixing of lipid and peptide. The final mass concentration of the stock solution was 0.5 mg/mL. For monolayer formation, 40 µL of either lipid or the lipid-aurein mixture was carefully spread onto the air-water interface using a glass Hamilton microsyringe.

The monolayer was allowed to equilibrate for 45 min to ensure complete evaporation of organic solvents. The temperature of the aqueous subphase was maintained at 37 °C using a circulating water bath. The monolayer was compressed symmetrically at a constant rate and the surface pressure was continuously monitored using the Wilhelmy plate method.

## RESULTS AND DISCUSSION

To investigate the effect of aurein on membrane mechanical properties, NSE measurements were performed on DMPC ULVs containing different concentrations of aurein . The ULVs were prepared by extrusion with careful optimization of the preparation protocol to minimize multilamellar structures, as described in detail in the Experimental Section. The vesicle morphology was independently assessed by SANS, and detailed analysis of the SANS data (Fig. S2) confirmed that the prepared samples are predominantly unilamellar. The SANS analysis also revealed a moderate decrease in bilayer thickness upon aurein incorporation (Table S1). Typically observed normalized intermediate scattering functions, $I(Q,t)/I(Q,0)$ for DMPC membrane with 2.5 mol % aurein at different $Q$ values are shown in Fig. 1(a). To highlight the dependence of aurein concentration, $I(Q,t)/I(Q,0)$ for DMPC membrane at different concentrations of aurein at a typical $Q$ value 0.09 Å$^{-1}$ are shown in Fig. 1 (b). It is evident that as the concentration of aurein or $P/L$ ratio increases, $I(Q,t)/I(Q,0)$ decays faster at the same $Q$ value suggesting that membrane becomes softer. To get microscopic insights, detailed analysis of intermediate scattering functions has been carried out.

Zilman and Granek (ZG) [35] have provided a formalism to describe the $I(Q,t)$ functions. ZG theory is applicable in the $Q$ range of $QR$>>1, where $R$ is the radius of vesicles. In this length scale, a membrane does not interact with its neighbouring membranes, hence the dynamic behaviour of the single membrane can be observed. According to ZG formulism, $I(Q,t)$ function of free standing single membrane due to thermal undulation motion decays with time as a stretched exponential with an exponent 2/3 and the relaxation rate increases with $Q^3$. In the present case, $R$ is ~500 Å (as observed using dynamic light scattering) and $Q_{\min}$ =0.048 Å$^{-1}$, which satisfy the $Q_{\min}R$ >>1 condition at all the $Q$ values. Hence, we have used ZG theory according to which the normalized intermediate scattering function for membranes can be written as[35]

$$\frac{I(Q,t)}{I(Q,0)} = \exp\left[-\left(\Gamma(Q)t\right)^{2/3}\right] \tag{1}$$

where $t$ is the Fourier time and $\Gamma(Q)$ is the $Q$-dependent relaxation rate for the bending motion and can be written as

$$\Gamma = 0.025\gamma\left(\frac{k_B T}{\kappa}\right)^{1/2}\frac{k_B T}{\eta}Q^3 \tag{2}$$

Here, $\kappa$ is the bending modulus of the membrane, $\eta$ is the solvent viscosity, $k_B$ is the Boltzmann's constant, and $T$ is the temperature. The factor $\gamma$ accounts for the orientational averaging between the membrane plaquettes and $\boldsymbol{Q}$ and approaches unity for $\kappa/k_B T >> 1$. For lipid bilayers this condition is well satisfied and thus $\gamma = 1$[20], [26], [27], [36].

Bending modulus of lipid membrane obtained using Eq. (2) was found to be higher by a factor of 9 compared to its expected value. This discrepancy arises from the assumptions inherent in the original ZG theory, which considers dissipation of membrane undulation energy solely through the surrounding solvent. The ZG model is based on the Helfrich bending Hamiltonian and does not account for internal membrane degrees of freedom, particularly neglecting intermonolayer friction within the bilayer. To address this limitation, Watson and Brown[37] extended the ZG framework by incorporating an additional dissipative mechanism arising from relative sliding (slipping mode) between the two monolayer leaflets. In this extended model, energy dissipation occurs through both the solvent and intrinsic membrane friction. While the relaxation rate retains its characteristic $Q^3$ dependence as in Eq. (2) the bending modulus $\kappa$ is replaced by an effective (or re-normalised) bending modulus $\tilde{\kappa}$. Hence, for lipid membrane in the extended ZG framework, Eq. (2) can be re-written as

$$\Gamma = 0.025\left(\frac{k_B T}{\tilde{\kappa}}\right)^{1/2}\frac{k_B T}{\eta}Q^3 \tag{3}$$

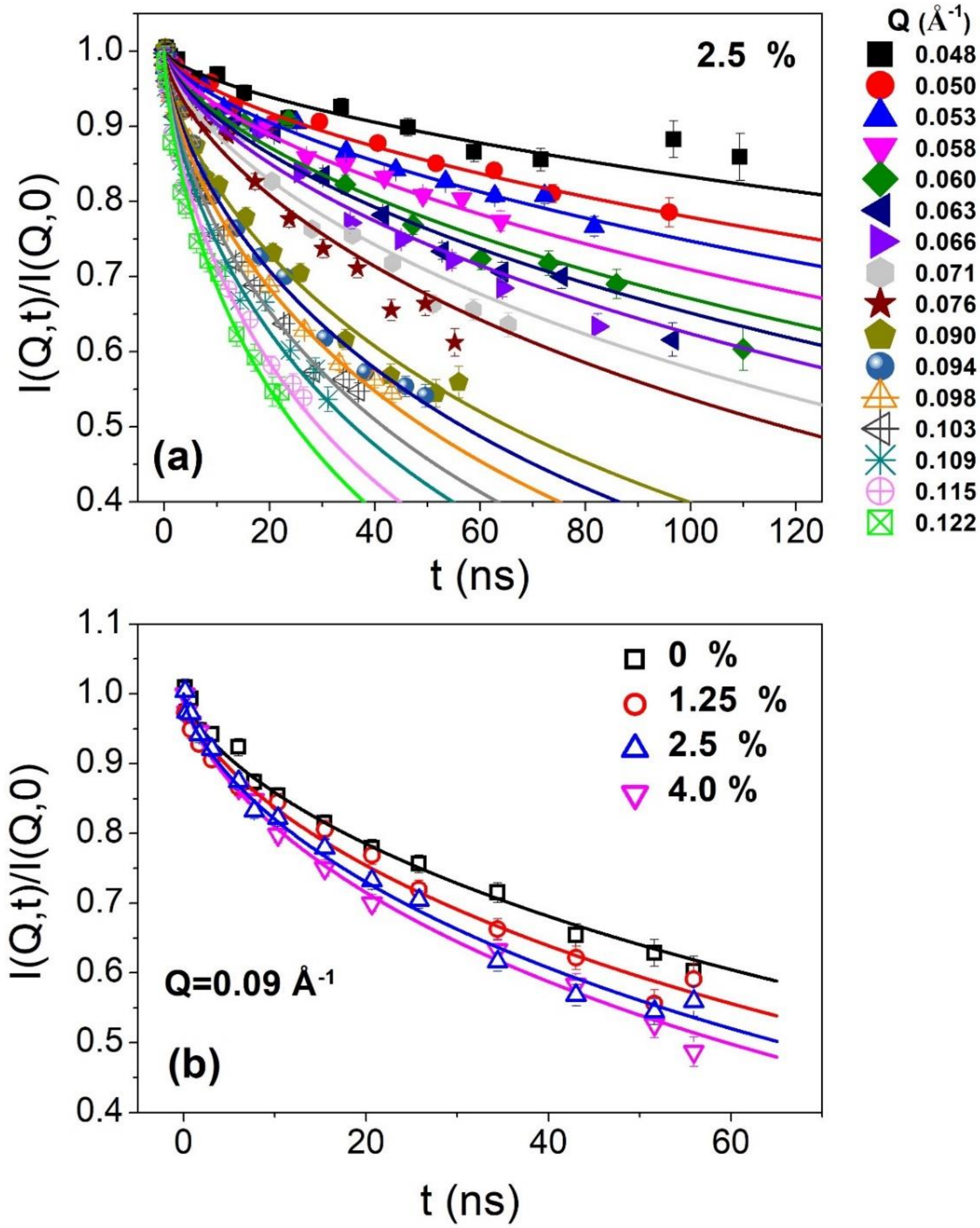

**Figure 1** Normalized intermediate scattering functions at 37°C :(a) DMPC membrane with 2.5 mol % aurein measured at different *Q* values , and (b) DMPC membrane at different concentrations of aurein at a particular $Q$=0.09Å$^{-1}$. Solid lines represent fits to the Zilman-Granek model ( Eq. (1)) capturing membrane bending dynamics.

Eq. (1) was used to describe the observed intermediate scattering function for DMPC membrane with different concentrations of aurein at 37°C. It is found that stretched exponential given by ZG theory describe the observed data well for DMPC membrane at all the concentrations of aurein at all the $Q$ values (Fig. 1 & Fig. S3). Typical fits are shown for DMPC membrane with aurein by the solid lines in Fig. 1(a) & (b). Obtained relaxation rates (Γ) are shown in Fig. 2 with $Q^3$ which follow a close to linear behaviour for all the systems. Eq. (3) is used to describe the behavior of the relaxation rate with $Q^3$, and the obtained values of slope are given in Table 1. It is found that the slope increases with increasing concentration of aurein suggesting faster undulation motion of the membrane. The re-normalised (or effective) bending modulus can be obtained from the slope as

$$\left(\frac{\tilde{\kappa}}{k_B T}\right)^{1/2} = \frac{0.345 \times T(K)}{m\left(\text{Å}^3/\text{ns}\right)\eta(cP)} \tag{4}$$

where $m$ is the slope between the relaxation rate Γ and $Q^3$. The re-normalised bending modulus $\tilde{\kappa}$ can be expressed in terms of intrinsic bending modulus of the membrane and compressibility modulus as[37]

$$\tilde{\kappa} = \kappa + 2d^2 K_m \tag{5}$$

Here $K_m$ is the monolayer compressibility modulus and $d$ is the height of the neutral surface from the bilayer midplane. The first component in Eq. (5) explains the contributions from the intrinsic bending modulus of membrane, κ, while the second term accounts for the dissipation inside the bilayer from the redistribution of the individual lipids. This term arises since membrane deformations are unable to relax in the finite time scale which leads to an increased energy penalty for deforming the membrane on the nanoscale. In presence of aurein, a membrane active peptide, there is no prior information about how bending modulus or $2d^2K_m$, i.e. dissipation within the bilayer is affected[38]. We therefore first focus on the relative change in the effective bending modulus ($\tilde{\kappa}$ or $\tilde{\kappa}/\tilde{\kappa}_0$) with respect to aurein concentration, here, $\tilde{\kappa}_0$ is the effective bending modulus of the membrane in absence of aurein ( i.e. $P/L$=0). This approach provides a direct and model-independent measure of change in effective bending modulus of the membrane without invoking specific assumptions regarding the relationship between κ and $\tilde{\kappa}$. At 37°C, viscosity of $D_2O$ is 0.84 cP[39]. Using Eq. (4), we determined the effective bending modulus $\tilde{\kappa}$ of DMPC membrane at different concentrations of aurein and obtained values are given in Table 1. It is evident that $\tilde{\kappa}$ monotonically decreases with increasing aurein concentration, indicating progressive membrane softening arising from peptide-induced disruption of lipid packing, which reduces the energetic cost of membrane curvature. Relative ratios $\tilde{\kappa}/\tilde{\kappa}_0$ shown in Fig. 3, decreases to ~0.47 at 4 mol% aurein, corresponding to more than a twofold reduction in bending rigidity.

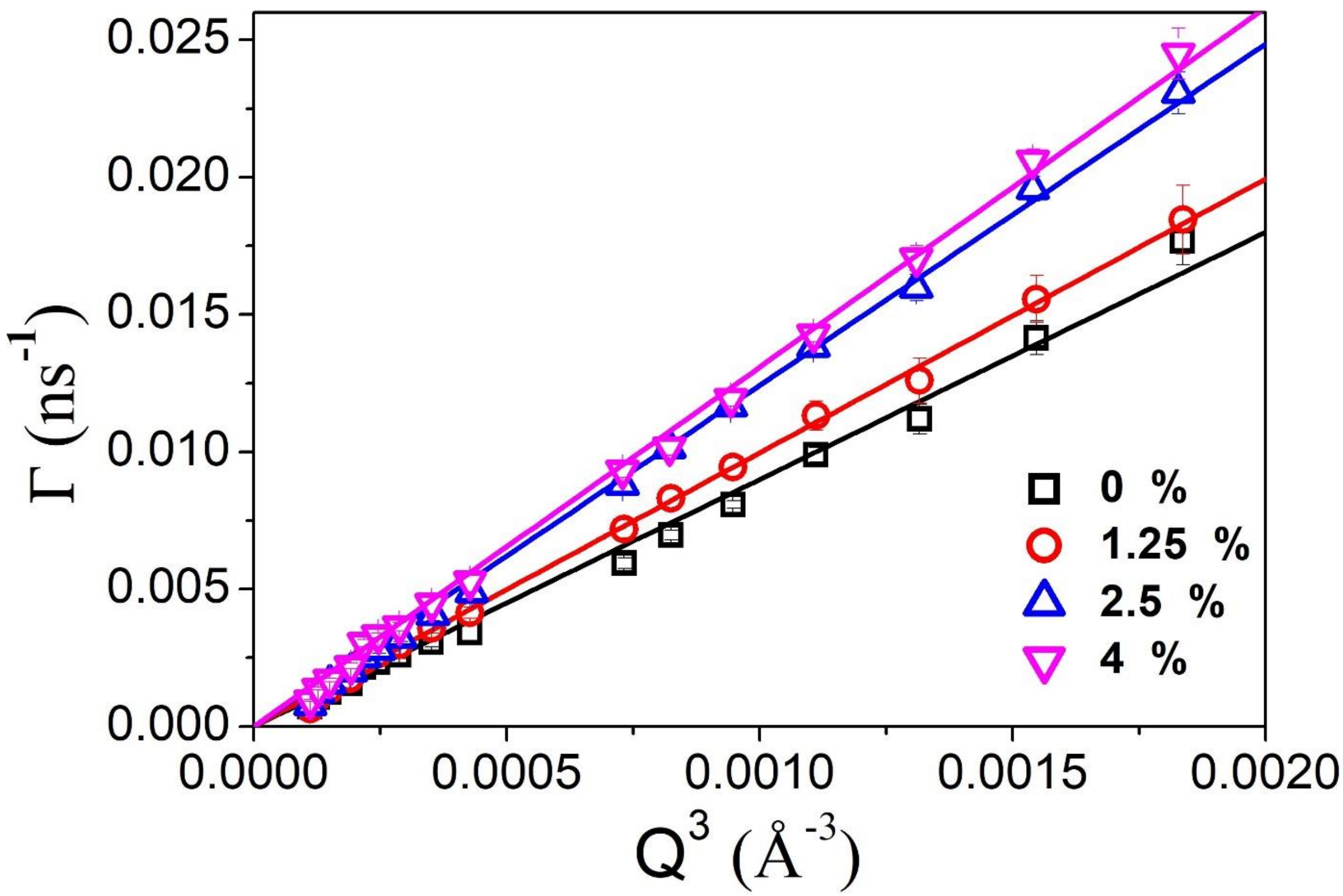

**Figure 2** Variation of relaxation rate (Γ) associated with membrane undulation dynamics as a function of $Q^3$ for DMPC membrane at different peptide to lipid ($P/L$) ratios. Solid lines represent least-squares fits using Eq. (3).

This monotonic softening contrasts with previously reported behavior in mixed zwitterionic–anionic membranes (DMPC/DMPG), where aurein induces an initial softening followed by pronounced stiffening at higher peptide concentrations, even in the absence of pore formation[18]. In the present study, spanning a wide concentration range up to $P/L$ = 4%, no such stiffening is observed. For comparison, the corresponding $\tilde{\kappa}/\tilde{\kappa}_0$ values reported previously for mixed DMPC/DMPG membranes (75:25 and 50:50) [18] are included in Fig. 3. This comparison is made in terms of nominal $P/L$ and is not intended as a thermodynamically normalized comparison at identical membrane-bound peptide concentration. Nevertheless, the qualitatively different concentration-dependent responses-progressive softening in DMPC versus non-monotonic softening followed by stiffening in DMPC/DMPG-demonstrate that membrane composition strongly influences the mechanical response to aurein.

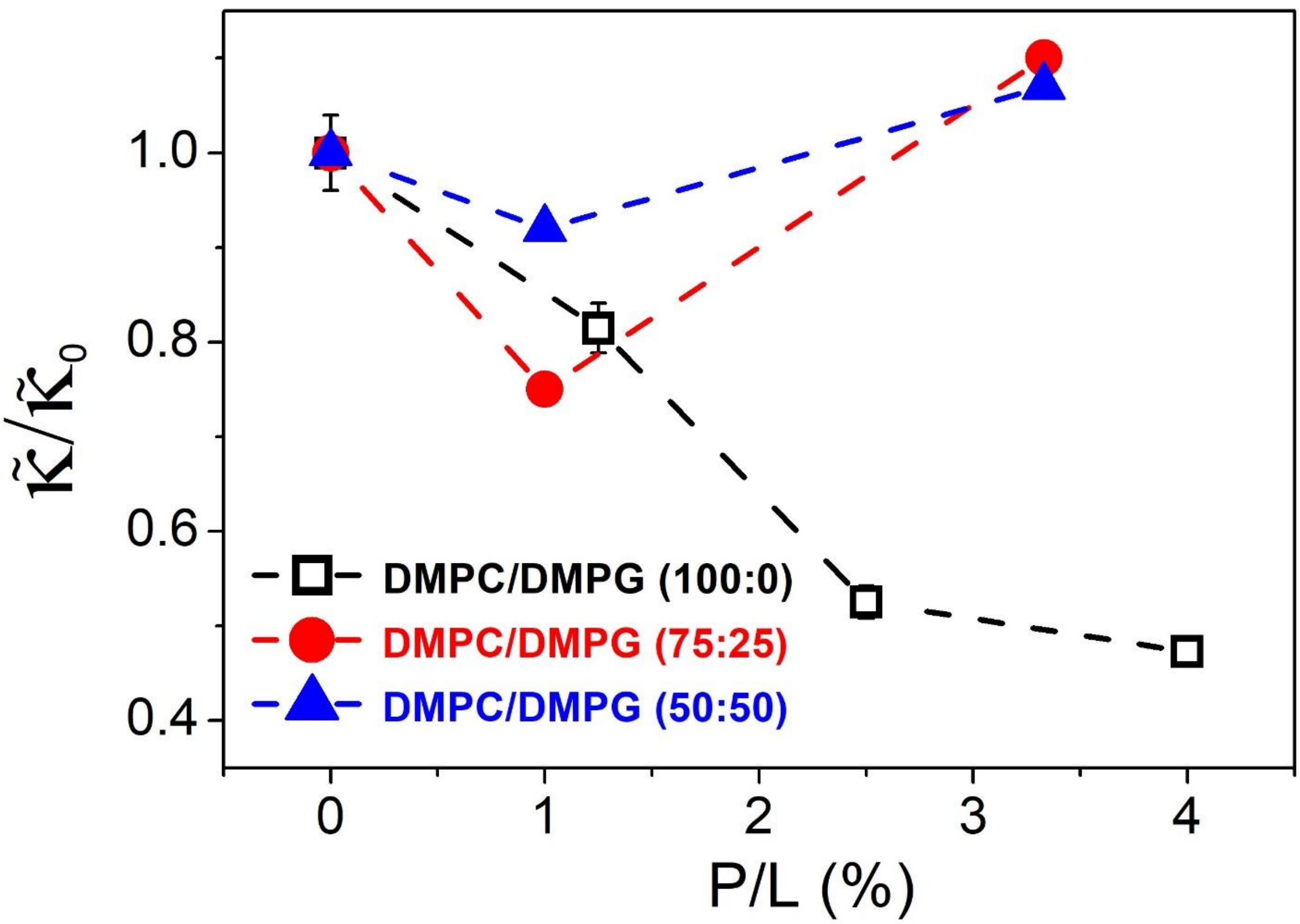

**Figure 3** Variation of relative ratio of effective bending moduli ($\tilde{\kappa}/\ \tilde{\kappa}_0$) of zwitterionic DMPC membrane as a function of nominal aurein P/L. A continuous decrease in $\tilde{\kappa}/\ \tilde{\kappa}_0$ with increasing peptide concentration indicates progressive membrane softening. For comparison, corresponding data for mixed zwitterionic and anionic lipid membrane (DMPC/DMPG; 75:25 and 50:50) [18] are included which exhibit a non-monotonic dependence on peptide concentration.

Intrinsic bending modulus κ can be estimated from $\tilde{\kappa}$ using Eq. (5) under certain approximations. Exact value of height of the neutral surface (*d*) cannot be observed experimentally. The neutral surface is defined as a surface monolayer that neither compresses nor stretches as it is bent. However, it should be around half of the bilayer thickness. Assuming a slab layer[40], compressibility modulus of monolayer $K_m$ can be expressed as

$$K_m = 12\frac{\kappa_m}{{d_m}^2} \tag{6}$$

where $\kappa_m$ is the monolayer bending modulus and $d_m$ is the monolayer hydrocarbon thickness. Assuming $\kappa_m = \kappa / 2$ and $d_m = d_t / 2$, where $d_t$ is hydrocarbon core thickness, Eq. (5) can be re-written as

$$\tilde{\kappa} = \left\{1 + 12(2d / d_t)^2\right\}\kappa \tag{7}$$

We assumed $2d/d_t$=1.0 similar to the earlier studies [24], [25] to keep the membrane neutral surface at the interface between the head group and tail rather than in headgroup region of the bilayer. This results in the effective bending modulus $\tilde{\kappa}$ =13 $\kappa$. Using this relation one can write Eq. (4) in terms of intrinsic bending modulus as

$$\left(\frac{\kappa}{k_B T}\right)^{1/2} = \frac{0.0957 \times T(K)}{m\left(\text{Å}^3/\text{ns}\right)\eta(cP)} \tag{8}$$

Values of intrinsic bending modulus has been calculated for DMPC membrane with different concentrations of aurein and given in Table 1. At 37°C, the bending modulus for neat DMPC membrane is obtained 15.2 ± 0.4 $k_B$T which is consistent with the literature [24], [25], [41]. Incorporation of aurein results in a monotonic decrease in the intrinsic bending modulus, reaching 7.2 ± 0.1 $k_B$T at 4 mol% aurein. This corresponds to an approximately 53% reduction, or a 2.1-fold decrease, in bending rigidity, demonstrating that aurein substantially softens the zwitterionic membrane. This pronounced softening of zwitterionic DMPC contrasts markedly with the non-monotonic behavior reported for mixed DMPC/DMPG membranes (75:25 and 50:50) [18]. For example, in the DMPC/DMPG (75:25) membrane, 1 mol% aurein produces a ~7% decrease in the intrinsic bending modulus, whereas at 3.3 mol% aurein, the bending modulus increases by ~8% relative to the peptide-free membrane, indicating an overall stiffening of the membrane at higher peptide concentration. Thus, while aurein progressively softens the zwitterionic DMPC membrane, its effect in anionic-containing membranes is distinctly non-monotonic, highlighting the critical role of membrane composition in determining the mechanical response to peptide incorporation.

**Table: 1 Average values of Γ/$Q^3$ (slope, *m*) obtained from Γ vs $Q^3$ plots, along with the effective bending modulus ($\tilde{\kappa}$), bending modulus (κ) and bilayer compressibility modulus ($K_A$) of DMPC membrane at different concentration of aurein at 37 °C**

| *P*/*L* (%) | *m* (Å$^3$/ns) | $\tilde{\kappa}$ ($k_B$T) | $\tilde{\kappa}/\tilde{\kappa}0$ | κ ($k_B$T) | $K_A$ (mN m$^{-1}$) |
|---|---|---|---|---|---|
| 0 | 9.00 ± 0.13 | 198 ± 6 | 1.0± 0.04 | 15.2 ± 0.4 | 216 ±10 |
| 1.25 | 9.97 ± 0.07 | 162 ± 2 | 0.81 ± 0.03 | 12.4 ± 0.2 | 180 ± 5 |
| 2.5 | 12.42 ± 0.10 | 104 ± 2 | 0.52 ± 0.02 | 8.0 ± 0.1 | 121 ± 6 |
| 4 | 13.09 ± 0.11 | 93 ± 1 | 0.47 ± 0.02 | 7.2± 0.1 | 110 ± 8 |

The contrasting mechanical responses of DMPC and DMPC/DMPG membranes to increasing aurein concentration raise an important structural question: whether the distinct mechanical pathways are associated with differences in the depth of peptide penetration into the bilayer. In particular, the pronounced, monotonic decrease in bending modulus observed for zwitterionic DMPC prompted us to examine whether the progressive softening could be linked to increasing penetration of aurein into the hydrophobic chain region. NMD provides detailed structural information on the spatial distribution of molecular components within the membrane, as well as the bilayer thickness [21], [22]. We therefore employed d54-DMPC, with deuterated acyl chains, together with protiated aurein to maximize the neutron-scattering contrast between the hydrocarbon region and the peptide. This contrast was specifically chosen to sensitively probe any substantial peptide-associated contribution within the deuterated chain region. Thus, the NMD measurements provide a direct structural test of whether the concentration-dependent decrease in bending modulus of zwitterionic DMPC membrane is accompanied by progressively deeper penetration of aurein into the hydrophobic membrane core. The molecular structure of d54-DMPC is shown in Fig. S1. The experiments were carried out at 30°C, above DMPC's phase transition temperature, at two different hydrations namely 94 % and 98% RH. All samples exhibited well-defined lamellar diffraction patterns with an oscillating rocking scan, with up to six orders of diffraction peaks observed. Typical observed two dimensional diffraction pattern for supported d54-DMPC membrane is shown in Fig. 4(a). This pattern is consistent with the expected form factor of a slab-like bilayer structure that is dominated by deuterated acyl chains (d54- DMPC). The corresponding one-dimensional diffraction profiles of DMPC membranes containing different concentrations of aurein are shown in Fig. 4(b). The diffraction data were corrected for various instrument effects such as pixel sensitivity, dark current, etc., normalized to the proton charge count, and then converted into a 2D map of reciprocal space vectors, *Q*r and *Q*z, using drt-SANS[42]. After these corrections, the diffraction peak intensities were integrated from the 2D position sensitive detector. The resulting integrated intensities were used to obtain the diffraction amplitudes (form factors) by applying standard corrections, including the Lorentz factor ($C_L$), absorption factor ($C_{abs}$), and geometry correction factor ($C_{geo}$). The form factors amplitudes $|F(h)|$ were calculated as

$$|F(h)| = \sqrt{\frac{I(h)}{I_0 C_L C_{abs} C_{geo}}} \tag{9}$$

where *h* is the Miller index. The fifth and sixth order are too weak due to the bilayer structure itself, therefore, first four orders of diffraction amplitude used for density reconstruction. Obtained $|F(h)\ (h=1, 2, 3, 4)|$ for DMPC with varying concentrations of aurein at 98 % RH are given in Table 2. Analysis of the hydration-dependent lamellar swelling (Fig. S4) provides an independent constraint on the bilayer structure and confirms the phase relationship of the bilayer components as (+1, -1, -1, -1). The form factors $|F(h)|$ and their phases were used to reconstruct the scattering length density (SLD) in real space by using a cosine Fourier sum.

The SLD along the bilayer normal, $\rho(x)$, was constructed from $F(h)$ by

$$\rho(x) = \sum_h \phi_h |F(h)| \cos\left(\frac{2\pi h x}{D}\right) \tag{10}$$

where $\phi_h$ is the phase factor, $D$ is the lamellar spacing (*d*-spacing) determined from the peak positions in the diffraction pattern. The SLD profiles were reconstructed for samples containing 0%, 1.25%, 2.5%, and 4% (molar ratio) aurein. These profiles were normalized using two reference points: the SLD of 8% $D_2O$ (0 fm/Å$^3$) at the bilayer edge and the SLD of $CD_2$ groups (0.7115 fm/Å$^3$) in the hydrocarbon core, the lowest and the highest SLDs of the bilayer, respectively. The reconstructed SLD profiles reveal subtle but important changes in the bilayer structure upon aurein binding (Figure 4c). The broad peak in the SLD profile corresponds to the deuterated acyl chains of d54-DMPC, with a dip at the bilayer center representing the region near terminal $CD_3$ groups, which is of lower SLD than $CD_2$ groups. As the peptide concentration increases, we observe a progressive shift in the density distribution, indicating a reorganization of the lipid molecules in response to peptide binding.

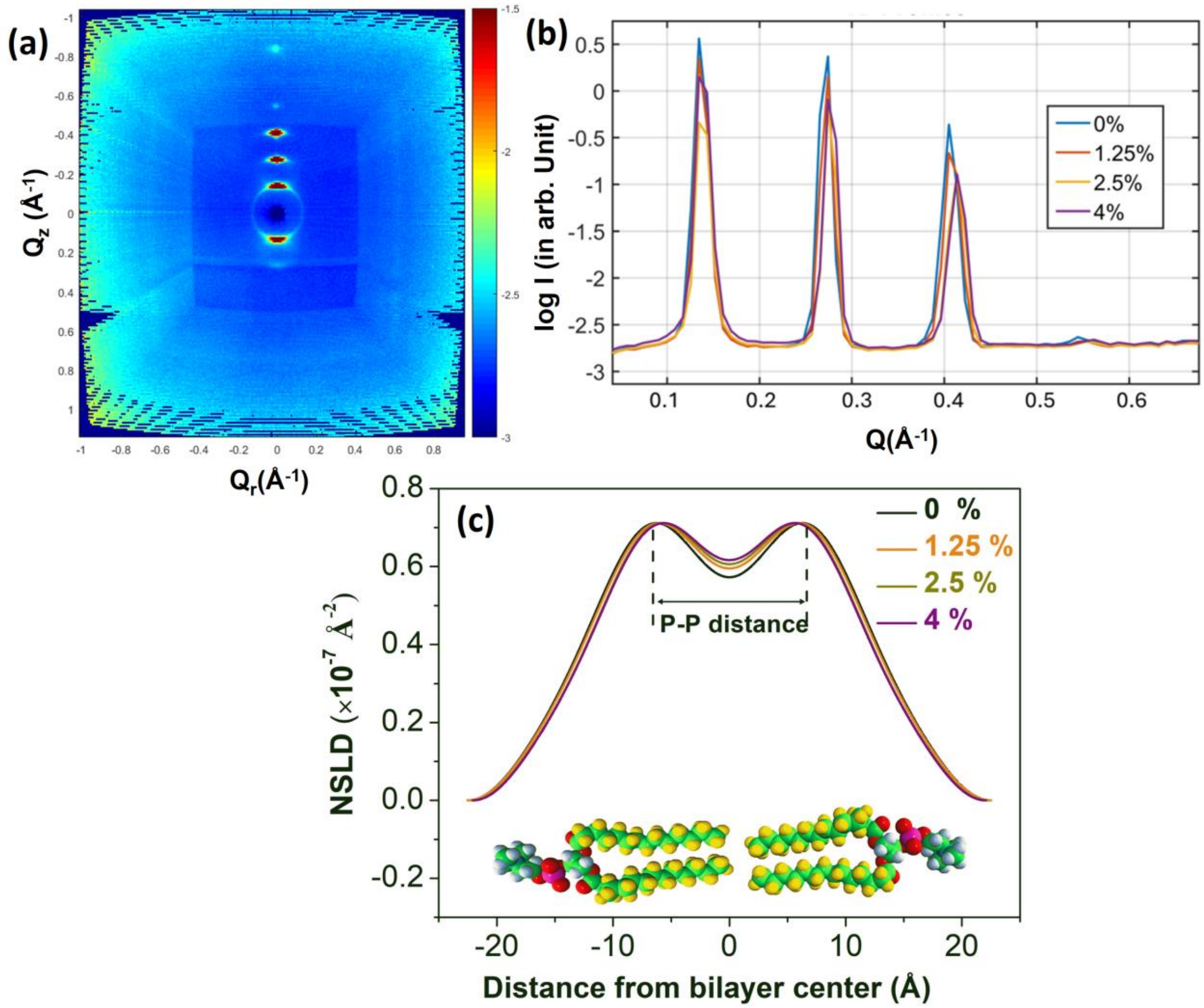

**Figure 4** Neutron membrane diffraction of DMPC bilayers in the presence of aurein. (a) Two-dimensional neutron diffraction pattern of DMPC at 98% RH, displayed on a logarithmic intensity scale. (b) One-dimensional diffraction profiles of DMPC membranes containing different concentrations of aurein at 98% RH (c) Neutron scattering length density (NSLD) profiles of DMPC bilayers without aurein (black), and with 1.25 mol % (orange), 2.5 mol % (yellow), and 4 mol % (purple) aurein as obtained using Eq. (10). The peak-to-peak distance between the $CD_2$ maxima, indicated by the dashed lines, represents the hydrocarbon-chain thickness. Two d54-DMPC molecular models are overlaid below the NSLD profiles to illustrate the approximate positions of the lipid headgroups and hydrocarbon chains. Deuterium and protium atoms are shown in yellow and white, respectively.

**Table 2 Form factors (arbitrary unit) and corresponding phases for first 4 diffraction peaks of DMPC membrane at varying aurein concentrations.**

| | | ***P/L* (%)** | | | |
|---|---|---|---|---|---|
| | **phase** | **0 %** | **1.25%** | **2.5%** | **4%** |
| ***F*(1)** | +1 | 7.50 | 7.65 | 7.23 | 7.74 |
| ***F*(2)** | −1 | 2.38 | 2.13 | 1.84 | 1.89 |
| ***F*(3)** | −1 | 1.13 | 1.05 | 0.93 | 0.93 |
| ***F*(4)** | −1 | 0.44 | 0.43 | 0.43 | 0.44 |

**Table 3 Structural parameters including lamellar spacing (*D*-spacing), peak-to-peak distance and hydrocarbon core thickness (FWHM of $CD_2$ peak) obtained from SLD profiles of DMPC bilayer as a function of aurein concentrations.**

| ***P/L*** | ***D*-spacing (Å)** | **Peak-to-Peak distance (Å)** | **Hydrocarbon core thickness ($d_t$) (FWHM) (Å)** |
|---|---|---|---|
| 0 % | 45.04 ± 0.15 | 12.57 ± 0.12 | 26.82 ± 0.10 |
| 1.25% | 45.04 ± 0.14 | 12.06 ± 0.10 | 26.56 ± 0.11 |
| 2.5% | 44.35 ± 0.14 | 11.52 ± 0.11 | 25.97 ± 0.13 |
| 4% | 44.13 ± 0.15 | 11.26 ± 0.10 | 25.82 ± 0.11 |

While the overall *d*-spacing of the bilayers shows only minor changes (Table 3), a more substantial effect is observed in the peak-to-peak (P-P) distance of the $CD_2$ maxima, which decreases from 12.57 Å in the absence of peptide to 11.26 Å at 4 mol % aurein. This reduction suggests a compression of the hydrophobic core upon peptide binding. Furthermore, the full width at half maximum (FWHM) of the broad $CD_2$ peak, which serves as an indicator of hydrocarbon chain thickness, decreases with increasing peptide concentration similarly. These observations collectively indicate that aurein binding induces a slight thinning of the DMPC bilayer.

Intriguingly, the SLD profiles show no abrupt changes or significant perturbations in the hydrocarbon region, which would be expected if aurein were inserted deeply into the bilayer core. The SLD of protiated peptide is substantially lower than that of the deuterated lipid chains (d54-DMPC), and any significant insertion would result in marked reduction to the SLD profile. The absence of such changes, even at 4 mol % peptide concentration, strongly suggests that aurein does not penetrate deeply into the hydrophobic core of DMPC bilayers. Instead, the SLD profiles are consistent with aurein remaining predominantly associated with the membrane interface, accompanied by a modest reduction in bilayer thickness. A subtle increase in the SLD near the bilayer center suggests a minor redistribution of the acyl chains, with $CD_2$ groups becoming more prevalent near the central $CD_3$ region. This effect could be attributed to the peptide "squeezing" the lipids, causing a minor reorganization of the hydrophobic core without direct insertion, consistent with the membrane thinning effect of membrane-active peptides[43], [44], [45], [46]. Thus, the pronounced monotonic decrease in bending rigidity observed by NSE cannot be attributed to progressively deeper penetration of aurein into the hydrophobic core. Rather, these results support an interfacial mode of peptide-membrane interaction in which aurein alters lipid packing and collective membrane elasticity without substantial concentration-dependent insertion into the hydrocarbon region. This behavior aligns well with previous studies on aurein and related AMPs[47], [48], which have been shown to exert their effects through membrane disruption mechanisms that do not necessarily involve deep penetration into the bilayer core.

In a continuum elastic description, the bending rigidity is related to both the bilayer thickness and the area compressibility modulus. Thus, membrane softening can arise from bilayer thinning, reduced resistance to in-plane area deformation, or a combination of both. Our NMD measurements indicate a modest aurein-induced thinning of the DMPC bilayer, suggesting that structural thinning alone may not fully account for the substantial reduction (about 53 %) in κ. We therefore investigated the bilayer area compressibility modulus, $K_A$ which can be obtained from the bending modulus of the membrane as[20]

$$K_A = 24\frac{\kappa}{d_t^{\,2}} \tag{11}$$

where $d_t$ is hydrocarbon core thickness and are given in Table 3. Using Eq. (11), the $K_A$ for DMPC at different concentrations of aurein were evaluated and the corresponding values are listed in Table 1. Our analysis reveals a clear and systematic decrease in $K_A$ with increasing aurein concentration, in striking contrast to the non-monotonic response observed for DMPC/DMPG[18], where $K_A$ initially decreases but subsequently increases, reaching a value above that of the peptide-free membrane at P/L=3.33 mol%. The increase in $K_A$ signifies enhanced resistance to in-plane deformation and therefore a recovery of membrane stiffness at higher aurein concentration. In contrast, the monotonic decrease in $K_A$ for DMPC indicates progressive weakening of lateral lipid–lipid cohesion without such cooperative stiffening. Importantly, the DMPC softening occurs without substantial changes in hydrocarbon-core thickness, indicating that the overall transbilayer architecture remains largely preserved. Within the elastic framework of Eq. (11), the pronounced reduction in κ can therefore be attributed predominantly to the weakening of in-plane area elasticity rather than to bilayer thinning. Thus, membrane composition dictates fundamentally different mechanical pathways: progressive loss of lateral cohesion and softening in zwitterionic DMPC versus concentration-dependent recovery of in-plane stiffness in anionic-containing DMPC/DMPG membranes.

To independently validate observed contrasting results in compressibility modulus, we performed $\pi$–$A$ isotherm measurements on DMPC and DMPC: DMPG (100:75) monolayer at the air-water interface at different concentrations of aurein. Observed isotherms for DMPC and DMPC/ DMPG (100:75) monolayers are shown in Fig. 5(a) and (b), respectively. The corresponding monolayer compressibility modulus $E_{CM} = -A\left(\frac{\partial \pi}{\partial A}\right)_T$ for both DMPC and DMPC/DMPG (100:75) monolayers as a function of surface pressure are presented in Fig. 5 (c) and (d), respectively. To enable a meaningful comparison between the bilayer area compressibility modulus ($K_A$) derived from NSE and monolayer compressibility modulus ($E_{CM}$) obtained from π–

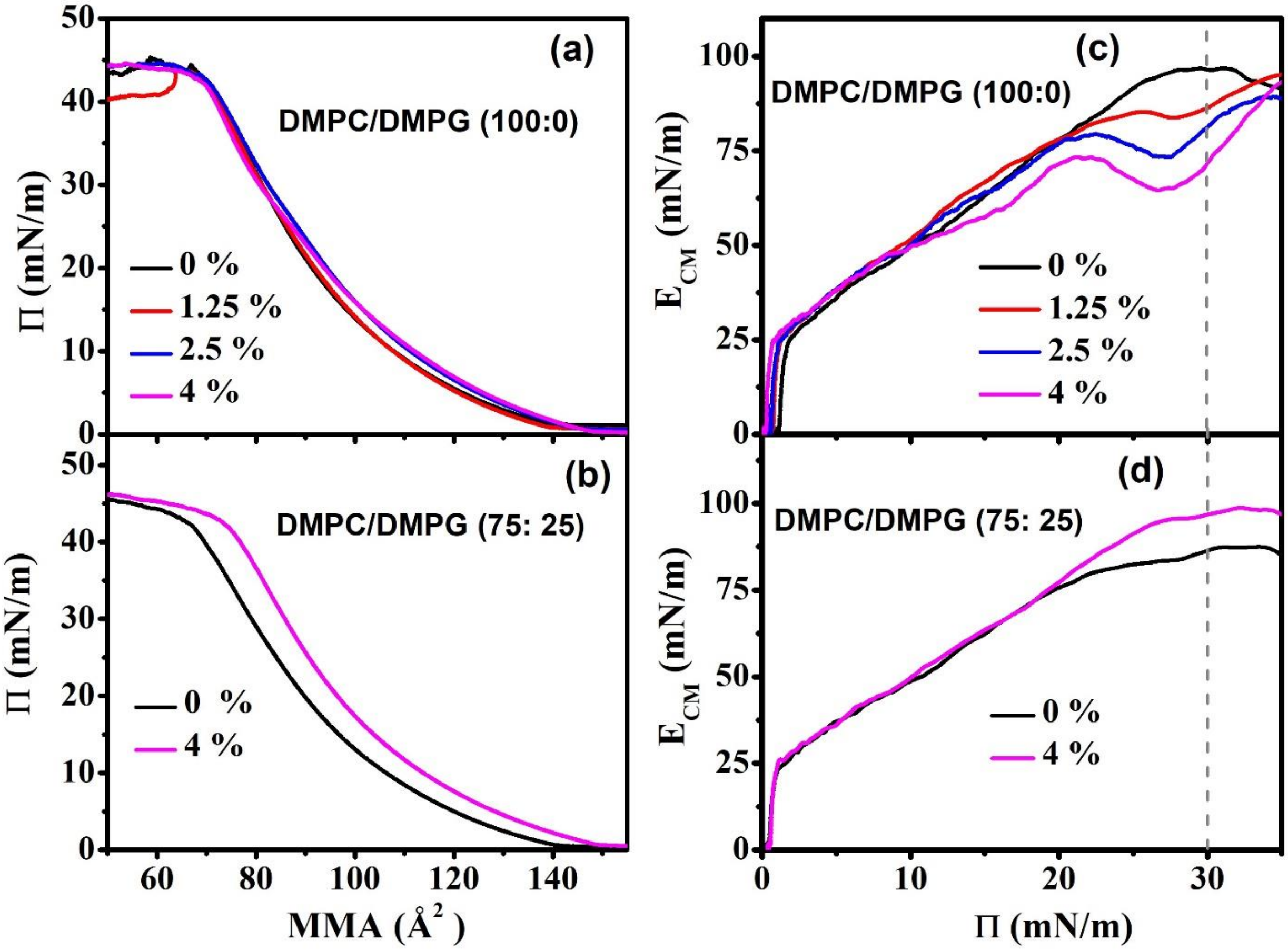

**Figure 5** Surface pressure (π)–area (A) isotherms of (a) DMPC and (b) DMPC/DMPG (100:75) monolayers formed at the air–water interface at 37 °C in the absence and presence of aurein. Corresponding monolayer compressibility moduli, $E_{CM}$, of (c) DMPC and (d) DMPC/DMPG (100:75) monolayers as a function of surface pressure at different aurein concentrations. The dashed line at 30 mN/m denotes the surface pressure corresponding approximately to the lipid packing conditions relevant to bilayer membranes. At this pressure, aurein produces contrasting changes in the compressibility modulus of these two lipid systems, highlighting the strong dependence of the membrane mechanical response on lipid composition.

A isotherms, the latter must be evaluated at a surface pressure of approximately 30 mN/m, which corresponds to the packing density of a fully hydrated bilayer[49], [50]. At this physiologically relevant packing regime, for DMPC monolayer, the π–*A* isotherms exhibit a clear decrease in in-plane monolayer compressibility modulus upon addition of aurein, in direct agreement with the NSE-derived reduction in bilayer compressibility modulus. In contrast, the DMPC/DMPG measurements reveal stiffening upon 4 mol% aurein incorporation, which is consistent with the corresponding NSE results. $\pi$–$A$ isotherm measurements provide independent and complementary validation of the composition-dependent mechanical responses observed by NSE, demonstrating that the distinct responses of zwitterionic DMPC and anionic-containing DMPC/DMPG are consistently reproduced by two complementary experimental approaches. Despite geometric differences, the consistent trends across these complementary approaches provide strong evidence that aurein weakens lateral lipid cohesion in zwitterionic DMPC without significantly altering bilayer thickness.

To place these findings in a broader compositional context, we compare NSE studies with mixed zwitterionic-anionic membranes (DMPC/DMPG), where the dependence of $\kappa$ on peptide concentration is non-monotonic [18]. In that system, $\kappa$ and $K_A$ initially decreases at low peptide-to-lipid ratios, consistent with membrane softening, but subsequently increases markedly at higher concentrations ($P/L \approx 3.33\%$), even exceeding the stiffness of the peptide-free membrane. The origin of this contrasting behavior can be understood in terms of electrostatic interactions and composition-dependent lateral organization. In purely zwitterionic DMPC membranes, aurein interacts primarily through amphipathic surface binding and shallow insertion, perturbing lipid packing and weakening lateral cohesive interactions. This reduces both $K_A$ and $\kappa$ in a continuous manner, consistent with a progressive reduction in collective elastic restoring forces without inducing large-scale structural heterogeneity. In contrast, mixed DMPC/DMPG membranes which contain anionic lipids, exhibit strong electrostatic attraction between cationic aurein and negatively charged PG headgroups. At elevated peptide concentrations, this promotes preferential binding and lateral segregation, leading to peptide-PG complexes enriched nanodomains[17]. The formation of such domains introduces lateral heterogeneity, thickness mismatch, and interfacial line tension within the bilayer. At low peptide concentrations, packing disruption dominates, resulting in membrane softening; however, beyond a threshold concentration, domain formation imposes elastic constraints that suppress long-wavelength undulations and increase the effective bending rigidity. In such heterogeneous membranes, the measured $\kappa$ reflects not only local molecular packing but also domain coupling and boundary energetics, resulting in an apparent stiffening at mesoscopic length scales probed by NSE. The absence of such non-monotonic behavior in pure

DMPC strongly suggests that lateral segregation is driven by electrostatic lipid-peptide interactions rather than generic amphipathic insertion. Zwitterionic membranes lack the charge contrast necessary for strong preferential binding and domain stabilization, leading instead to uniform softening without mesoscale stiffening. A schematic illustration of these contrasting behaviors is presented in Fig. 6. This distinction establishes a composition-dependent mechanical switch governing aurein-membrane interactions.

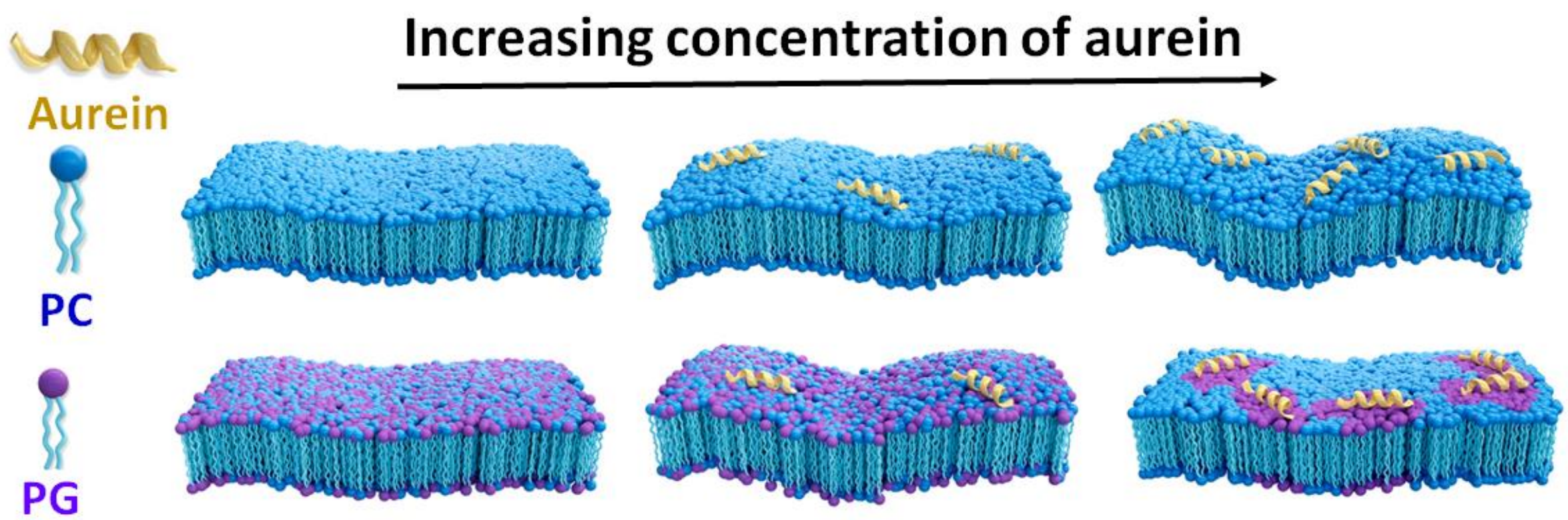

**Figure 6** Schematic representation of the composition-dependent mechanical response of membranes to aurein. Zwitterionic PC bilayers show monotonic softening with increasing peptide concentration, reflecting uniform weakening of lipid cohesion. In contrast, mixed PC/ PG bilayers display a non-monotonic behavior, where initial softening at low peptide concentration transitions to membrane stiffening at higher concentrations due to peptide-induced lateral lipid segregation and nanodomain formation.

A particularly key finding of this study emerges from combining the present NSE measurements with our earlier QENS results on the same DMPC-aurein system[17]. Aurein induces a remarkable decoupling between molecular lipid dynamics and collective membrane mechanics: lipid lateral diffusion decreases monotonically with increasing aurein concentration, while $\kappa$ and $K_A$ decrease concurrently. Thus, aurein restricts local lipid motion while simultaneously softening the membrane at the collective mesoscopic scale. This demonstrates that suppression of molecular lipid mobility does not necessarily imply enhanced membrane rigidity. The distinct responses arise because lateral diffusion is governed primarily by local friction, transient peptide-lipid interactions, and dynamic heterogeneity [51], whereas κ and $K_A$ describe collective restoring forces and lipid packing. We propose that aurein adsorption disrupts collective lipid packing and weakens membrane elasticity, while transient peptide-lipid interactions and increased interfacial friction impede local lipid transport. This multiscale decoupling provides an important physical insight into how peptide binding can simultaneously constrain local molecular lipid motion and promote collective membrane softening. It demonstrates that molecular lipid dynamics and collective membrane elasticity can exhibit contrasting responses to the same peptide perturbation, representing distinct manifestations of peptide-membrane interactions that emerge at different length and time scales.

## CONCLUSIONS

Antimicrobial peptide (AMP)-membrane interactions are strongly influenced by membrane composition and its associated physical properties. Here, we demonstrate that the prototypical AMP aurein induces progressive elastic softening of DMPC bilayers, a zwitterionic PC-rich model membrane, manifested by monotonic decreases in both the bending rigidity (κ) and area compressibility modulus ($K_A$). This softening is accompanied by reduced lateral lipid diffusion at the molecular scale[17], revealing a striking decoupling between collective membrane elasticity and molecular lipid dynamics. Thus, aurein can simultaneously restrict molecular lipid motion while weakening the collective elastic resistance of the membrane, highlighting distinct physical responses of the membrane across length and time scales.

In contrast, anionic phosphatidylglycerol (PG)-containing DMPC/DMPG membranes exhibit a non-monotonic mechanical response, with initial softening followed by pronounced stiffening at higher aurein concentrations, consistent with peptide-induced lipid segregation and nanodomain formation[18]. The absence of this high-concentration stiffening in zwitterionic DMPC demonstrates that membrane composition strongly influences the mechanical pathway through which aurein perturbs the membrane.

Together, these findings establish a direct connection between peptide-lipid interactions, molecular lipid dynamics, and collective membrane mechanics, demonstrating that membrane composition can qualitatively alter the physical response of membranes to aurein. The results provide a testable physical framework for investigating composition-dependent AMP-membrane interactions in increasingly complex and physiologically representative membrane systems, while recognizing that membrane composition, membrane architecture, and additional cellular components collectively contribute to AMP function in biological environments.

**ACKNOWLEDGEMENTS**

This research used resources at the Spallation Neutron Source (SNS), a DOE Office of Science User Facility operated by the Oak Ridge National Laboratory. Beam times were allocated on EQ-SANS and NSE spectrometers on proposal IPTS–33346.1 and 15683, respectively. We thank Dr. L. Stingaciu for assistance with the NSE data acquisition at SNS.

**DATA AVAILABILITY**

Unless otherwise stated, all data supporting the results of this study can be found in the article, supplementary, and source data files. Experimental data can be accessed through Zenodo under DOI: 10.14461/oncat.data/3674459.

**CODE AVAILABILITY**

The code used to analyze the NMD data is available from the authors upon reasonable request. The function used for the analysis of the SANS data can be created as the sum of two core_multi_shell functions of SasView and is also available from the authors upon reasonable request.

## REFERENCES

[1] M. Naghavi, S. E. Vollset, K. S. Ikuta, L. R. Swetschinski, A. P. Gray, E. E. Wool, G. Robles Aguilar, T. Mestrovic, G. Smith, C. Han, et al.,, *The Lancet* **2024**, 404, 1199.

[2] N. Mookherjee, M. A. Anderson, H. P. Haagsman, D. J. Davidson, *Nature Reviews Drug Discovery* **2020**, 19, 311.

[3] M. Magana, M. Pushpanathan, A. L. Santos, L. Leanse, M. Fernandez, A. Ioannidis, M. A. Giulianotti, Y. Apidianakis, S. Bradfute, A. L. Ferguson, A. Cherkasov, M. N. Seleem, C. Pinilla, C. de la Fuente-Nunez, T. Lazaridis, T. Dai, R. A. Houghten, R. E. W. Hancock, G. P. Tegos, *The Lancet Infectious Diseases* **2020**, 20, e216.

[4] J. B. Mitra, V. K. Sharma, M. Kumar, A. Mukherjee, *Mini Reviews in Medicinal Chemistry* **2020**, 20, 183.

[5] B. H. Gan, J. Gaynord, S. M. Rowe, T. Deingruber, D. R. Spring, *Chemical Society Reviews* **2021**, 50, 7820.

[6] M. Zasloff, *Nature* **2002**, 415, 389.

[7] V. Teixeira, M. J. Feio, M. Bastos, *Progress in Lipid Research* **2012**, 51, 149.

[8] V. K. Sharma, E. Mamontov, D. B. Anunciado, H. O'Neill, V. S. Urban, *Soft Matter* **2015**, 11, 6755.

[9] V. K. Sharma, E. Mamontov, M. Tyagi, S. Qian, D. K. Rai, V. S. Urban, *The Journal of Physical Chemistry Letters* **2016**, 7, 2394.

[10] J. Bhatt Mitra, V. K. Sharma, A. Mukherjee, V. Garcia Sakai, A. Dash, M. Kumar, *Langmuir* **2020**, 36, 397.

[11] S. Raghav, P. Hitaishi, R. P. Giri, A. Mukherjee, V. K. Sharma, S. K. Ghosh, *Journal of Materials Chemistry B* **2024**, 12, 11731.

[12] D. Kaushik, V. K. Sharma, S. K. Ghosh, in *Advances in Biomembranes and Lipid Self-Assembly*, Vol. 41 (Eds: A. Iglič, M. Rappolt, P. Losada-Pérez), Academic Press, 2025.

[13] S. Raghav, B. Pati, N. Jaglan, A. Mukherjee, V. K. Sharma, A. Bandyopadhyay, S. K. Ghosh, *Soft Matter* **2025**, 21, 8692.

[14] J. B. Mitra, V. K. Sharma, M. Kumar, V. G. Sakai, A. Mukherjee, *The Journal of Physical Chemistry Letters* **2026**, 17, 4442

[15] T. Rozek, K. L. Wegener, J. H. Bowie, I. N. Olver, J. A. Carver, J. C. Wallace, M. J. Tyler, *European Journal of Biochemistry* **2000**, 267, 5330.

[16] D. K. Rai, S. Qian, *Scientific Reports* **2017**, 7, 3719.

[17] V. K. Sharma, S. Qian, *Langmuir* **2019**, 35, 4152.

[18] S. Qian, P. A. Zolnierczuk, *BBA Advances* **2022**, 2, 100045.

[19] S. Qian, V. K. Sharma, L. A. Clifton, *Langmuir* **2020**, 36, 15189.

[20] V. K. Sharma, E. Mamontov, *Progress in Lipid Research* **2022**, 87, 101179.

[21] S. Qian, D. K. Rai, *The Journal of Physical Chemistry Letters* **2018**, 9, 5778.

[22] S. Qian, G. Nagy, P. Zolnierczuk, E. Mamontov, R. Standaert, *The Journal of Physical Chemistry Letters* **2024**, 15, 4745.

[23] V. K. Sharma, E. Mamontov, M. Ohl, M. Tyagi, Phys. Chem. Chem. Phys. **2017**, 19, 2514.

[24] J. D. Nickels, X. Cheng, B. Mostofian, C. Stanley, B. Lindner, F. A. Heberle, S. Perticaroli, M. Feygenson, T. Egami, R. F. Standaert, J. C. Smith, D. A. Myles, M. Ohl and J. Katsaras, *J. Am. Chem. Soc*., **2015**, 137, 15772–15780.

[24] V. K. Sharma, M. Nagao, D. K. Rai, E. Mamontov, *Physical Chemistry Chemical Physics* **2019**, 21, 20211.

[25] M. Nagao, E. G. Kelley, R. Ashkar, R. Bradbury, P. D. Butler, *The Journal of Physical Chemistry Letters* **2017**, 8, 4679.

[26] S. Chakraborty, M. Doktorova, T. R. Molugu, F. A. Heberle, H. L. Scott, B. Dzikovski, M. Nagao, L.-R. Stingaciu, R. F. Standaert, F. N. Barrera, J. Katsaras, G. Khelashvili, M. F. Brown, R. Ashkar, *Proceedings of the National Academy of Sciences* **2020**, 117, 21896.

[28] A. Seth, S. K. Ghosh, V. K. Sharma, in *Advances in Biomembranes and Lipid Self-Assembly*, Vol. 42 (Eds: A. Iglič, M. Rappolt, P. Losada-Pérez), Academic Press, 2025.

[29] H. L. Scott, A. Skinkle, E. G. Kelley, M. N. Waxham, I. Levental, F. A. Heberle, *Biophysical Lett* **2019**, 17 , 1381

[30] T. Kaasgaarda, O. G. Mouritsenb, K. Jørgensen, *Biochimica et Biophysica Acta* **2003**, 1615 , 77– 83

[31] W.T. Heller, V. S. Urban, G. W. Lynn, K. L. Weiss, H. M. O'Neill, S. V. Pingali, S. Qian, et al., *Applied Crystallography* **2014**, 47 (4), 1238-1246

[32] J. K. Zhao, C. Y. Gao, D. Liu, *Journal of Applied Crystallography* **2010**, 43, 1068.

[33] M. J. M. Darkes, J. P. Bradshaw, *Acta Crystallogr. D* **2000**, 56 , 48–54

[34] M. Ohl, M. Monkenbusch, N. Arend, T. Kozielewski, G. Vehres, C. Tiemann, M. Butzek, H. Soltner, U. Giesen, R. Achten, H. Stelzer, B. Lindenau, A. Budwig, H. Kleines, M. Drochner, P. Kaemmerling, M. Wagener, R. Möller, E. B. Iverson, M. Sharp, D. Richter, *Nuclear Instruments and Methods in Physics Research Section A: Accelerators, Spectrometers, Detectors and Associated Equipment* **2012**, 696, 85.

[35] A. G. Zilman, R. Granek, *Physical Review Letters* **1996**, 77, 4788.

[36] T. Kumarage, S. Gupta, N. B. Morris, F. T. Doole, H. L. Scott, L.-R. Stingaciu, S. V. Pingali, J. Katsaras, G. Khelashvili, M. Doktorova, M. F. Brown, R. Ashkar, *Nature Communications* **2025**, 16, 7024.

[37] M. C. Watson, Frank L. H. Brown, *Biophysical Journal* **2010**, 98, L9.

[38] E. G. Kelley, P. D. Butler, M. Nagao, *Soft Matter* **2021**, 17, 5671.

[39] R. C. Hardy, R. L. Cottington, *Journal of Research of the National Bureau of Standards* **1949**, 42, 573.

[40] D. H. Boal, *Mechanics of the Cell*, Cambridge University Press, Cambridge **2012**.

[41] S. Gupta, J. U. De Mel, G. J. Schneider, *Current Opinion in Colloid & Interface Science* **2019**, 42, 121.

[42] W. T. Heller, J. Hetrick, J. Bilheux, J. M. B. Calvo, W.-R. Chen, L. DeBeer-Schmitt, C. Do, M. Doucet, M. R. Fitzsimmons, W. F. Godoy, G. E. Granroth, S. Hahn, L. He, F. Islam, J. Lin, K. C. Littrell, M. McDonnell, J. McGaha, P. F. Peterson, S. V. Pingali, S. Qian, A. T. Savici, Y. Shang, C. B. Stanley, V. S. Urban, R. E. Whitfield, C. Zhang, W. Zhou, J. J. Billings, M. J. Cuneo, R. M. F. Leal, T. Wang, B. Wu, *SoftwareX* **2022**, 19, 101101.

[43] M.-T. Lee, F.-Y. Chen, H. W. Huang, *Biochemistry* **2004**, 43, 3590.

[44] S. Qian, W. T. Heller, *The Journal of Physical Chemistry B* **2011**, 115, 9831.

[45] D. K. Rai, S. Qian, W. T. Heller, *Biochimica et Biophysica Acta (BBA) - Biomembranes* **2016**, 1858, 2788.

[46] H. W. Huang, *Biochemistry* **2000**, 39, 8347.

[47] A. Mechler, S. Praporski, K. Atmuri, M. Boland, F. Separovic, L. L. Martin, *Biophysical Journal* **2007**, 93, 3907.

[48] T.-H. Lee, C. Heng, M. J. Swann, J. D. Gehman, F. Separovic, M.-I. Aguilar, *Biochimica et Biophysica Acta (BBA) - Biomembranes* **2010**, 1798, 1977.

[49] D. Marsh, *Biochimica et Biophysica Acta (BBA) - Reviews on Biomembranes* **1996**, 1286, 183.
[50] A. Blume, *Biochimica et Biophysica Acta (BBA) - Biomembranes* **1979**, 557, 32.
[51] V. K. Sharma, H. Srinivasan, J. Gupta, S. Mitra, *Soft Matter* **2024**, 20, 7763.